\documentclass[showpacs, amsmath,
reprint,
aps]{revtex4-1}

\usepackage{graphicx}

\begin{document} 

\title{Giant circular dichroism in individual carbon nanotubes induced by extrinsic chirality} 

\author{A.~Yokoyama}
\author{M.~Yoshida}
\author{A.~Ishii}
\author{Y.~K.~Kato}
\email[e-mail: ]{ykato@sogo.t.u-tokyo.ac.jp}
\affiliation{Institute of Engineering Innovation, 
The University of Tokyo, Tokyo 113-8656, Japan}

\begin{abstract}
Circular dichroism is widely used for characterizing organic and biological materials, but measurements at a single molecule level are challenging because differences in absorption for opposite helicities are small. Here we show that extrinsic chirality can induce giant circular dichroism in individual carbon nanotubes, with degree of polarization reaching 65\%. The signal has a large dependence on the incidence angle, consistent with the interpretation that mirror symmetry breaking by the optical wave vector is responsible for the effect. We propose that field-induced charge distribution results in an efficient polarization conversion, giving rise to the giant dichroism. Our results highlight the possibility of polarization manipulation at the nanoscale for applications in integrated photonics and novel metamaterial designs.
\end{abstract}


\maketitle 

Materials with mirror symmetry breaking exhibit differences in absorption for the two helicities of circular polarization, which is known as circular dichroism (CD). It is a fundamental effect in electromagnetism, and yet it finds broad applications in analytical chemistry, molecular biology, and crystallography for identification and characterization of chiral molecules and solids \cite{Hecht, Berova}. Despite its common use on macroscopic samples, detection of CD from single molecules is nontrivial as the difference in absorption is typically less than a part in a thousand. The use of superchiral optical fields can enhance enantioselectivity \cite{Hendry:2010, Tang:2011}, but CD spectroscopy at a single molecule level still remains a challenge  \cite{Hassey:2006, Tang:2009, Barnes:2009, Cohen:2009, Cyphersmith:2012}. In contrast to the smallness of the effect in molecules, giant CD at optical frequencies has been demonstrated in metamaterials \cite{Gonokami:2005, Gansel:2009, Plum:2009, Soukoulis:2011}. Such strong chirality effects are expected to open up novel opportunities in optics and photonics including negative index media \cite{Pendry:2004}, but carefully designed metal nanostructures with appropriate resonances are required.

Here we report on the observation of giant CD in individual single-walled carbon nanotubes (CNTs), and show that it arises from extrinsic chirality \cite{Plum:2009}. Spectrally and spatially resolved CD measurements are performed  on single air-suspended CNTs using photoluminescence (PL) for detection, unambiguously revealing circularly polarized absorption originating from the nanotubes. The degree of polarization reaches a value as high as 65\%, an unforeseen level of CD in single nanoparticles. Surprisingly, we find that the CD signal is strongly dependent on the angle of incidence, changing its sign and vanishing at certain angles. All nanotubes investigated have shown angle-dependent polarization with the same sign, against the expectation that half of the nanotubes show the opposite sign if the nanotube handedness is responsible for the dichroism. Rather, the results are consistent with CD that is induced by extrinsic chirality, where the optical wave vector breaks the mirror symmetry under oblique incidence. We propose a microscopic physical mechanism in which spatially-distributed field-induced charge on the substrate surface causes enhancement or reduction of absorption for circular polarization. The results demonstrate the feasibility of polarization manipulation using extrinsic chirality at the level of single nanoparticle, opening up the possibility of polarization engineering at the nanoscale. 

Our samples are single-walled carbon nanotubes suspended over trenches on silicon substrates. Electron beam lithography and dry etching processes are performed to form the $\sim$5~$\mu$m deep trenches on (001) Si wafers. Another electron beam lithography step defines the catalyst windows, and nanotubes are grown by chemical vapor deposition using ethanol as a carbon source \cite{Maruyama:2002, Imamura:2013}. The nanotubes are then characterized with a laser scanning confocal microspectroscopy system \cite{Moritsubo:2010, Yasukochi:2011, Watahiki:2012}. Wavelength-tunable continuous-wave Ti:sapphire laser is used for excitation, and an objective lens with a numerical aperture of 0.7 and a working distance of 10~mm focuses the laser and collects the PL at a resolution of $\sim$1.5~$\mu$m. An InGaAs photodiode array attached to a spectrometer is used for obtaining PL spectra. A steering mirror scans the laser spot to image and identify well isolated CNTs, while chiral indices are assigned by PL excitation spectroscopy \cite{Bachilo:2002, Ohno:2006prb}. PL intensity dependence on linear polarization angle allows the determination of the nanotube axis.

\begin{figure}
\includegraphics{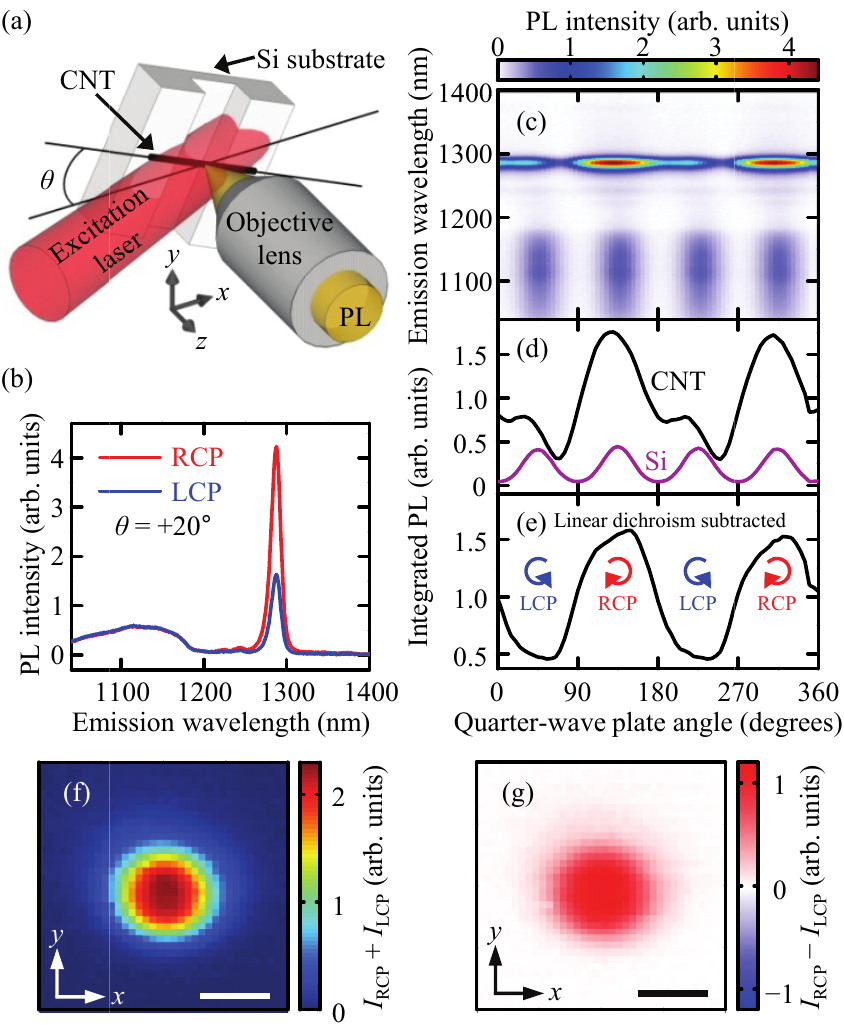}
\caption{\label{fig1}
(a) A schematic of the experimental geometry. The substrate is parallel to the $xy$ plane and the optical collection path through the objective lens is along the $z$-axis.
(b) PL spectra of an individual nanotube for excitation with RCP (red) and LCP (blue). 
(c) PL spectra as a function of the quarter-wave plate angle. Laser is linearly polarized along the $y$-axis before the wave plate.
(d) Integrated PL intensity from CNT (black) and Si (purple) as a function of the quarter-wave plate angle. The spectral integration window for the CNT is centered at 1290~nm and has a width of 40~nm, while Si PL has been integrated from 1040~nm to 1200~nm.
(e) Linear dichroism for CNT PL is extracted by fitting the data in (d) by two sinusoidal functions with periods of 90$^\circ$ and 180$^\circ$, and the component with 90$^\circ$ periodicity has been subtracted.
(f) and (g) Sum and difference images, respectively, of integrated CNT PL for excitation with RCP and LCP. The excitation laser spot is fixed and the collection spot is scanned to obtain the images. Scale bars are 2~$\mu$m. For (b-g), $\lambda=780$~nm and $P=26$~mW are used for excitation. The nanotube is at $\theta=+20^\circ$ and has a suspended length of 1.3~$\mu$m. 
}\end{figure}

After we identify and characterize individual CNTs, CD measurements are performed in the side excitation geometry as shown in Fig.~\ref{fig1}(a). The excitation laser beam is directed at a glancing angle of $\sim$10$^\circ$ from the sample plane, lightly focused by a lens with a focal length of 75~mm to an elliptical spot with full-widths at half-maximum of $\sim$60~$\mu$m and $\sim$20~$\mu$m along the $x$ and $y$ directions, respectively. The laser is initially linearly polarized along the $y$-axis, and an achromatic quarter-wave plate is placed just before the lens to obtain circular polarization. The sample is mounted on an automated rotation and three-dimensional translation stage to allow in-plane incidence-angle dependence measurements. All measurements are done at room temperature in air.

Figure~\ref{fig1}(b) shows typical PL spectra taken at an in-plane angle $\theta=+20^\circ$ for excitation with right-circular polarization (RCP) and left-circular polarization (LCP). The excitation wavelength $\lambda$ is tuned to the $E_{22}$ resonance of this $(9,7)$ tube, and a power $P=26$~mW is used. The sharp peak near 1300~nm is the emission from the CNT, while Si PL is the weaker luminescence centered around 1100~nm. It is clear that the nanotube PL is modulated by a factor of $\sim$2.5 for RCP and LCP.

Such a change in PL intensity can arise from beam deviation with wave plate rotation, but we use an actuator driven mirror to correct the deviation of the laser spot to below 2~$\mu$m. The fact that Si PL intensity is nearly the same for the two helicities shows that the excitation intensity changes are insignificant, and therefore the change in CNT PL must arise from laser polarization.

As CNTs have strong linearly polarized absorption \cite{Hartschuh:2003, Lefebvre:2004, Moritsubo:2010}, inaccuracies in the angle of the quarter-wave plate can give rise to an artifact. To distinguish linear and circular dichroism signals, we have performed the measurements as a function of the quarter-wave plate angle [Fig.~\ref{fig1}(c)]. Linear dichroism gives rise to intensity modulation with a 90$^\circ$ period, as in the case for Si PL where $p$-polarization has a larger transmission coefficient and thus results in stronger PL [Fig.~\ref{fig1}(d)]. In comparison, the nanotube PL intensity modulation has a component with a period of 180$^\circ$ [Fig.~\ref{fig1}(e)], a signature of circularly polarized absorption. 

As another possible source of an artifact, we have examined the polarization of the reflected laser. If the polarization is significantly different for RCP and LCP after reflection, and if the reflected laser is partly responsible for the excitation of CNTs, differences in PL intensity for incoming RCP and LCP could occur. We dismiss such an interpretation, because we have observed that the reflected polarization is largely $s$-polarized for both helicities as expected for glancing reflection from a substrate with a large index of refraction.

Next, we show that the signal is localized at the nanotube position by using PL images under circularly polarized excitation. We define $I_{\text{RCP}}$ and $I_{\text{LCP}}$ to be integrated PL intensity for excitation with RCP and LCP, respectively. We compute the sum $I_{\text{RCP}}+I_{\text{LCP}}$ and the difference $I_{\text{RCP}}-I_{\text{LCP}}$ to construct the images shown in Fig.~\ref{fig1}(f) and (g), respectively. These images show that the position and the extent of the difference signal overlaps with that of the sum signal. As the observed helicity-dependent intensity modulation is both spectrally and spatially localized at the nanotube,  we interpret the modulation in PL intensity as arising from CD of the nanotube.

\begin{figure}
\includegraphics{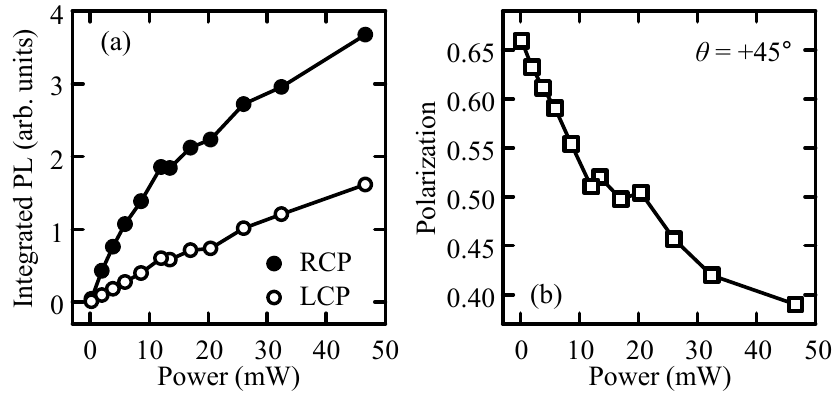}
\caption{\label{fig2}
(a) PL intensity as a function of the laser power for RCP (solid circles) and LCP (open circles), measured at $\theta=+45^\circ$ and $\lambda=780$~nm for a CNT with suspended length of 0.9~$\mu$m.  
(b) Excitation power dependence of the CD polarization calculated from data in (a).
}\end{figure}

A quite significant aspect of the CD signal we observe is its magnitude. We define the polarization $\rho=(I_{\text{RCP}}-I_{\text{LCP}})/(I_{\text{RCP}}+I_{\text{LCP}})$, and for the data shown in Fig.~\ref{fig1}(b), we obtain $\rho=45$\%. Furthermore, we find that the apparent polarization increases as the excitation intensity is lowered. On another $(9,7)$ tube, we have investigated the excitation power dependence (Fig.~\ref{fig2}). The polarization reaches a value as high as $\rho=65$\% at the lowest power that we used. Such a power dependent behavior is, however, somewhat expected as the PL intensity in CNTs increases sublinearly with excitation power \cite{Wang:2004, Xiao:2010, Moritsubo:2010}. The PL efficiency is reduced at high excitation rates because of exciton-exciton annihilation, and this is clearly observed for the case of RCP in Fig.~\ref{fig2}(a).  For LCP, such a sublinear behavior is negligible because the absorption is weaker. The reduction in PL efficiency for RCP results in smaller difference between PL intensities from what one would expect from the difference in absorption. The apparent reduction of polarization at higher excitation powers is therefore not attributed to changes in the CD of CNTs, and data at low excitation powers are more representative of the actual absorption polarization.

\begin{figure}
\includegraphics{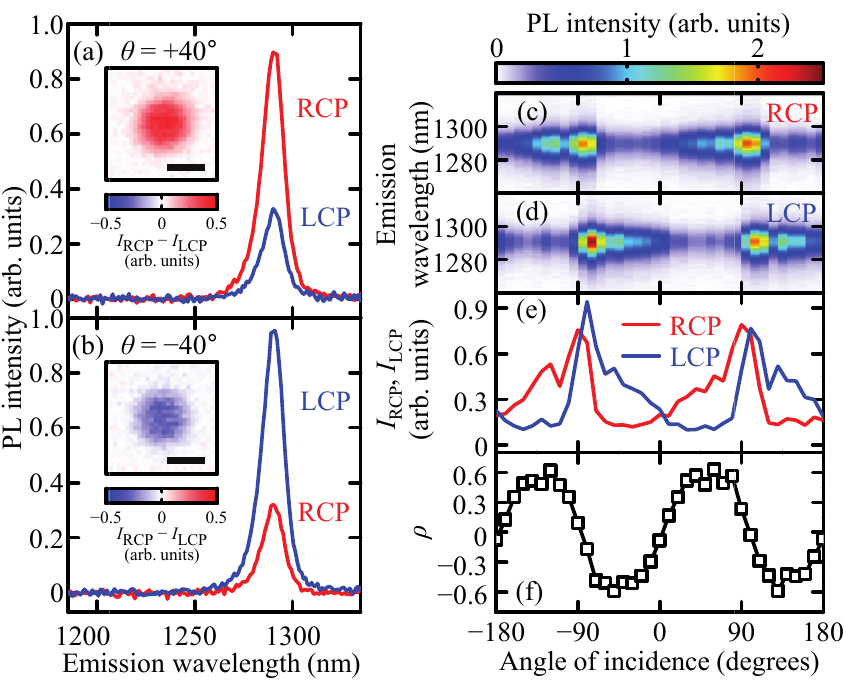}
\caption{\label{fig3}
(a) and (b) PL spectra taken at  $\theta=+40^\circ$ and $-40^\circ$, respectively, with RCP (red) and LCP (blue). Insets show the difference images. Scale bars are 2~$\mu$m. 
(c) and (d) In-plane angle dependence of PL spectra for RCP and LCP, respectively.
(e) $I_{\text{RCP}}$ (red) and $I_{\text{LCP}}$ (blue) as a function of $\theta$.
(f) Polarization calculated from data in (e).
Data are taken at $\lambda=780$~nm and $P=2$~mW on a 0.9~$\mu$m long tube.
}\end{figure}

Although it is known that chiral CNTs lack mirror symmetry and can exhibit CD after separation of enantiomers \cite{Peng:2007, Ghosh:2010}, the polarization that we observe is about three orders of magnitude larger. The fact that the CD signal is vastly larger compared to those observed in ensemble measurements already suggests that the origin of CD may be different from the intrinsic chirality of CNTs. It turns out that geometrical considerations shed light on the underlying mechanism.

Figure~\ref{fig3}(a) and (b) show PL spectra taken at two different in-plane incidence angles of $\theta=+40^\circ$ and $-40^\circ$, respectively. Strikingly, the polarization is inverted for these two geometries. To study such a behavior in more detail, we have performed measurements over a full rotation of the nanotube [Fig.~\ref{fig3}(c) and (d)]. The integrated PL is maximized at $\theta=\pm 90^\circ$ for both RCP and LCP as expected from strong absorption for linear polarization along the tube axis, but the PL intensities are similar for the two helicities [Fig.~\ref{fig3}(e)]. At oblique angles, however, PL intensities are different, and in fact we find that the polarization is strongly dependent on $\theta$ [Fig.~\ref{fig3}(f)]. In particular, the signal vanishes when the in-plane optical axis is parallel or perpendicular to the nanotube axis ($\theta=0^\circ, \pm 90^\circ, \pm 180^\circ$).

Such geometry dependent polarization suggests that extrinsic chirality \cite{Plum:2009} plays an important role, in which the optical wave vector $\vec{k}$ breaks the mirror symmetry. If we ignore the presence of the trench, a straight CNT on a substrate has two mirror symmetry planes perpendicular to the substrate, where one is normal to the CNT and the other includes the CNT. When the optical axis is outside of these two planes, the mirror symmetry is broken and extrinsic chirality emerges. This symmetry argument is consistent with the angle dependence of the observed polarization.

\begin{figure}[t]
\includegraphics{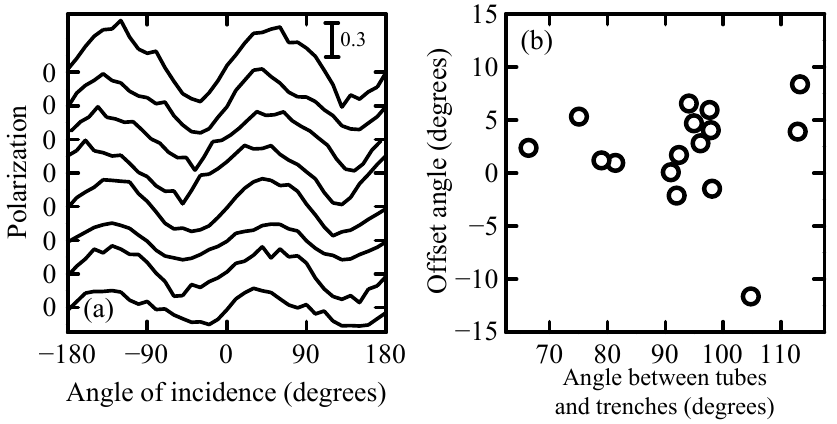}
\caption{\label{fig4}
(a) Angle dependent polarization for eight representative $(9,7)$ CNTs. The data are offset for clarity.
(b) Offset angle plotted as a function of tube angle with respect to the trenches. Sixteen tubes with a chirality (9,7) are used to generate this plot.
}\end{figure}

Statistics also support the interpretation that extrinsic chirality induces the CD. As we expect equal amounts of right- and left-handed CNTs in our samples, the sign of the polarization should be different for half of the tubes if the polarization originates from intrinsic chirality. Figure~\ref{fig4}(a) shows the angle dependent polarization for eight CNTs, where all tubes show positive polarization at $\theta=+45^\circ$ and negative polarization at $\theta=-45^\circ$. Such a behavior is reproduced in all 25 tubes that we have investigated, including those with different chiral indices. 

We note that all tubes in Fig.~\ref{fig4}(a) show similar angle dependence, although the angle between these tubes and the trenches range from $66^\circ$ to $113^\circ$. In order to look for any effects of the angle of the trenches, we fit the incidence angle dependent data to $\rho=A+B \sin [2(\theta-\theta_0)]$ where $A$ and $B$ are constants and $\theta_0$ is the offset angle. In Fig.~\ref{fig4}(b), we plot $\theta_0$ as a function of the angle between the nanotubes and the trenches, but we do not see a strong correlation. This result shows that the incidence angle with respect to the trench does not play a decisive role in the observed CD. 

\begin{figure}[t]
\includegraphics{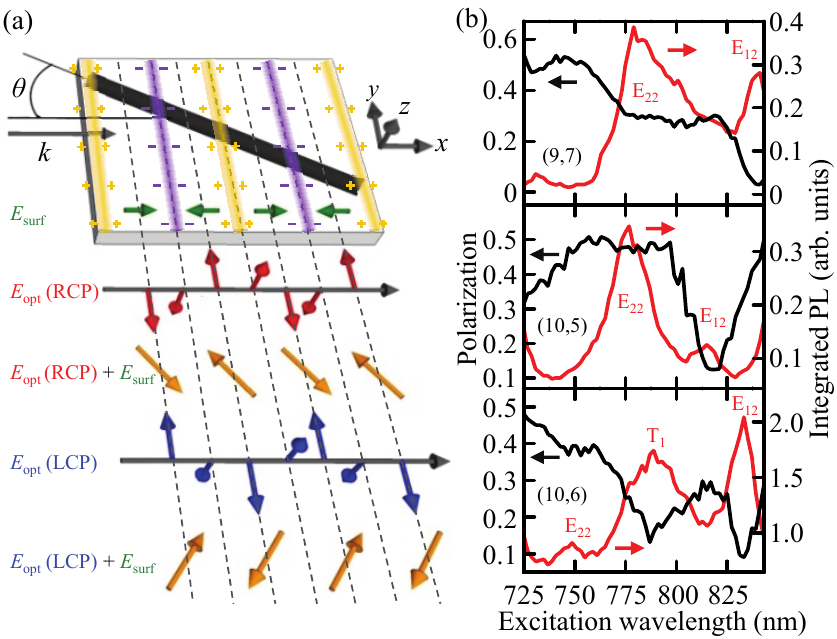}
\caption{\label{fig5}
(a) A schematic of the model for giant CD as explained in the text.
(b) Black curves show CD spectra taken at $\theta=+45^\circ$. The excitation powers ranges from 9~mW to 28~mW during wavelength tuning. Red curves show PL excitation spectra taken under perpendicularly polarized excitation at normal incidence. Top, middle, and bottom panels correspond to 2.5-$\mu$m-long $(9,7)$ tube, 3.0-$\mu$m-long $(10,5)$ tube, and 2.4-$\mu$m-long $(10,6)$ tube, respectively. The excitation spectra are normalized to powers of 170, 40, and 110~$\mu$W for the $(9,7)$, $(10,5)$, and $(10,6)$ tubes, respectively. The peaks are assigned according to Ref. \cite{Lefebvre:2007}.
}\end{figure}

Now we propose a microscopic model that explains our observations based on extrinsic chirality. Figure~\ref{fig5}(a) shows a schematic of the model. We assume the excitation laser path to be along the $x$-axis, ignore the presence of the trench, and consider the case for $0^\circ < \theta < +90^\circ$. We do not take into account the intrinsic chirality of the CNT, but use the fact that linear polarization  parallel to the tube axis dominates in the absorption process \cite{Hartschuh:2003, Lefebvre:2004, Moritsubo:2010}.  In Fig.~\ref{fig5}(a), a snapshot of the spatially varying optical electric field $\vec{E}_{\text{opt}}$ is shown for both helicities. At positions where $\vec{E}_{\text{opt}}$ is normal to the substrate, there will be induced charge on the surface which generates in-plane surface electric fields $\vec{E}_{\text{surf}}$ parallel to $x$. We note that $\vec{E}_{\text{surf}}$ does not depend on the helicity of excitation.

At positions where $\vec{E}_{\text{opt}}$ is parallel to the substrate, the sum $\vec{E}_{\text{opt}} + \vec{E}_{\text{surf}}$ can have considerably different orientations for the two helicities. For RCP, the total field will tend to be oriented at a smaller angle to the nanotube axis, while for LCP they tend to be oriented more perpendicular to the nanotube axis. Essentially, the presence of the substrate converts the circular polarization to in-plane linear polarization, enhancing or reducing the absorption depending on the helicity. Such a conversion is distinctly different from conventional means as the resultant polarization correspond to a different direction of the optical wave vector. 

This model can qualitatively explain the angle dependence of the CD. Inverting the angle will swap the roles of RCP and LCP, and the polarization changes sign. At $\theta=0^\circ$ and $\pm90^\circ$, the total field for RCP and LCP becomes symmetric with respect to the nanotube axis, and therefore  the CD vanishes.
In addition, CD spectra are also consistent with the proposed model.  As linearly polarized absorption parallel to the tubes is required in this model, the effect is expected to weaken at the $E_{12}$ transition where the absorption  polarization is reduced \cite{Lefebvre:2007, Miyauchi:2010}. Indeed, we observe broadband polarization with some decrease near $E_{12}$ transitions [Fig.~\ref{fig5}(b)]. Such a wavelength dependence is in contrast to ensemble measurements which have shown CD peaks at $E_{22}$ and $E_{33}$ transitions \cite{Peng:2007, Ghosh:2010}. 

\begin{figure}[t]
\includegraphics{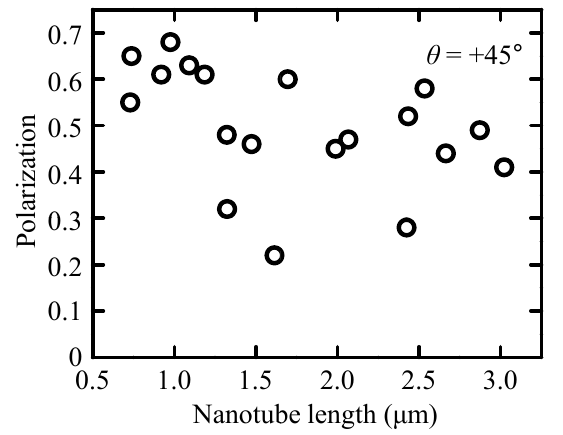}
\caption{\label{fig6}
Dependence of CD polarization on suspended length of (9,7) CNTs. Data are taken with $\theta = +45^\circ$,  $\lambda=780$~nm and $P=5$~mW.
}\end{figure}

We have also investigated suspended length dependence of the polarization. As the presence of the substrate is the essential origin of the effect in the model, the polarization is expected to lower as the suspended length of the CNT increases. The suspended part does not have the substrate underneath, and locally it possesses mirror symmetry so it should contribute less to CD. Such a dependence is more or less observed in Fig.~\ref{fig6}, but against our expectation, the polarization remains for lengths much longer than a wavelength. Further investigation is necessary to clarify this weak dependence.

It is unexpected that extrinsic chirality can induce such a giant response in a system as simple as a nanotube lying on a substrate. In principle, arrays of nanotubes on waveguides can be used to filter circularly polarized light or to convert polarization on a chip. Such a polarization conversion should also occur for semiconductor nanowires that have linearly polarized absorption. The proposed mechanism only relies on the electric response, which is fundamentally different from conventional understanding that the mixing of electric and magnetic response is needed for CD \cite{Plum:2009}. We hope that our finding will lead to new techniques for utilizing extrinsic chirality for polarization manipulation at the nanoscale.

\begin{acknowledgments}
We thank K. Konishi and H. Tamaru for helpful discussions, and acknowledge support from KAKENHI (21684016, 23104704, 24340066, 24654084), SCOPE, Asahi Glass Foundation,  KDDI Foundation, and the Photon Frontier Network Program of MEXT, Japan. The samples were fabricated at the Center for Nano Lithography \& Analysis at The University of Tokyo.
\end{acknowledgments}

\bibliography{GCDinCNT}

\end{document}